\documentclass[aps,prl,twocolumn,floatfix,showpacs,amssymb]{revtex4-1}
\usepackage[utf8]{inputenc}
\usepackage[english]{babel}
\usepackage{amsmath}
\usepackage{amsfonts}
\usepackage{amssymb}
\usepackage{graphicx}
\usepackage{float}
\usepackage{color}
\usepackage{braket}
\usepackage{mciteplus}
\usepackage[breaklinks=true,colorlinks,citecolor=blue,linkcolor=blue,urlcolor=blue]{hyperref}

\newcommand{\be}{\begin{equation}}
\newcommand{\ee}{\end{equation}}
\newcommand{\ber}{\begin{eqnarray}}
\newcommand{\eer}{\end{eqnarray}}

\def\nablabold{\mbox{\boldmath $\nabla$}}

\newcommand{\pv}{{\bf p}}
\newcommand{\Ev}{{\bf E}}

\newcommand{\Dv}{{\bf D}}

\newcommand{\Iv}{{\bf I}}
\newcommand{\jv}{{\bf j}}
\newcommand{\rv}{{\bf r}}
\newcommand{\qv}{{\bf q}}
\newcommand{\kv}{{\bf k}}

\newcommand{\vv}{{\bf v}}

\def\rhov{\mbox{\boldmath $\rho$}}
\def\eer{\end{eqnarray}}

\def\rhov{\mbox{\boldmath $\rho$}}

\def\nablabold{\mbox{\boldmath $\nabla$}}
\def\sigmabold{\mbox{\boldmath $\sigma$}}
\def\rv{{\bf r}}

\def\pv{{\bf p}}

\def\vv{{\bf v}}

\def\uv{{\bf u}}
\def\xv{\hat {\bf x}}
\def\jv{{\bf j}}
\def\kv{{\bf k}}
\def\qv{{\bf q}}

\def\Ev{{\bf E}}

\def\uv{{\bf u}}
\def\vv{{\bf v}}

\begin{document}

\title{Breakdown of the Wiedemann-Franz law in AB-stacked bilayer graphene}

\author{Mohammad Zarenia$^1$  and Giovanni Vignale$^{1,2}$}
\affiliation{$^1$Department of Physics and Astronomy, University of Missouri, Columbia, Missouri 65211, USA\\
$^2$Yale-NUS College, 16 College Ave West, 138527 Singapore}
\author{Thomas Benjamin Smith and Alessandro Principi}
\affiliation{School of Physics, University of Manchester, Oxford Road, Manchester M13 9PL,
UK}

\begin{abstract}
We present a simple theory of thermoelectric transport in bilayer graphene and report our results for the electrical resistivity,
the thermal resistivity, the Seebeck coefficient, and the Wiedemann-Franz ratio as functions of doping density and temperature. In the absence of disorder, the thermal resistivity tends to zero as the charge neutrality point is approached; the electric resistivity jumps from zero to an intrinsic finite value, and the Seebeck coefficient diverges in the same limit. Even though these results are similar to those obtained for single-layer graphene, their derivation is considerably more delicate.
The singularities are removed by the inclusion of a small amount of disorder, which leads  to the appearance of a ``window" of doping densities $0<n<n_c$ (with $n_c$ tending to zero in the zero-disorder limit) in which the Wiedemann-Franz law is severely violated.
\end{abstract}
\maketitle
{\it Introduction -- } The electric and thermal transport properties of graphene-based devices are a topic of great interest.  
Even setting aside their great potential for real-world applications, these systems have already offered unprecedented opportunities
to study new modalities of transport, in which hydrodynamic flow patterns, governed by global conservation laws, supersede the conventional diffusive dynamics of individual carriers \cite{narozhnyRev,Principi_prb_2016,Narozhny_prb_2015,Briskot_prb_2015,lucasReview}.  The simultaneous presence of carriers of opposite polarities --  electrons and holes -- whose density can be tuned by chemical doping, electrostatic gating, or simply by changing the temperature, creates a rich scenario of transport behaviors \cite{bandurin,ghahari,kumar,Gurzhi}. A particularly interesting one has recently been observed in a single layer of ultra-clean graphene near the charge neutrality point (CNP), where the chemical potential $\mu=0$.  The system is a zero-gap semiconductor with linearly dispersing conduction and valence bands and equal numbers of electrons and holes arising from thermal fluctuations at finite temperature.     Because $\mu=0$ the thermal (entropy) current coincides with the energy current,  and because of the linear dispersion the latter coincides with the total momentum density, which is a constant of the motion as long as impurities, lattice vibrations and umklapp effects are negligible. Thus we have an interesting situation in which the thermal resistivity vanishes while the electric resistivity remains finite because electrons and holes, moving in opposite directions under the action of an electric field, exert mutual friction on each other \cite{zarenia,Fritz_2008,Muller_2008,svintsov2012}. The result is that the Wiedemann-Franz (WF) ratio between the electric resistivity and the thermal resistivity is enhanced well above the standard value of $L_0 \equiv \pi^2(k_{\rm B}/e)^2/3$ (the so-called Wiedemann-Franz law) -- an effect that is indeed observed experimentally  \cite{crossno}, but only in a narrow window of doping densities around the CNP -- a window whose width shrinks to zero as the system is made less and less disordered. It should also be noted that this behavior is diametrically opposite to what one expects and observes in heavily doped graphene: in that case, only one polarity of carriers contributes to both electric and thermal transport and the electric resistivity plummets (barring electron-impurity and electron-phonon scattering)  while the thermal resistivity rises to a finite value:  in this case, the WF ratio drops {\it below} $L_0$ \cite{Principi_conductivity,lucasWF,lucasKT,foster}.

Motivated by these interesting findings, in this paper we investigate thermoelectric transport in a more complex system,  AB-stacked bilayer graphene (BLG), which is a zero-gap semiconductor with electron-hole symmetric parabolic bands touching at the Dirac point \cite{McCann}.  BLG is a quite distinct system from its single layer with the possibility of having a gate-induced tunable band gap \cite{Zhang2009}.  We ask in particular, whether a large violation of the WF law will still be present in the double-layer system near CNP.  At first glance,  a major qualitative difference exists between the two systems, because neither the energy current nor the particle current are conserved in the bilayer.  There seems to be no reason why the thermal resistivity would vanish in a bilayer at CNP when momentum-non-conserving processes are negligible.  Indeed, a naive calculation, based on the textbook theory of thermoelectric transport, would lead precisely to this conclusion: that the thermo-electric transport coefficients are free of singularities and qualitatively similar to what was obtained in single-layer graphene after the inclusion of disorder. 

One of the main purposes of this paper is to show that the naive conclusion is, in fact, incorrect.  After introducing a more careful treatment of the Boltzmann equation, which includes a conserved mode in which electrons and holes travel in the same direction, we are able to show that the transport coefficients remain singular as long as total momentum is conserved: that is to say, we find that, just as in single-layer graphene \cite{zarenia}, the electrical resistivity jumps from zero to a finite universal value (controlled by the strength of the Coulomb interaction) at the CNP, the thermal resistivity tends to zero, and the Seebeck coefficient diverges.  This happens because the conserved mode -- electrons and holes traveling in the same direction -- strongly overlaps the energy current mode and provides a dissipation-free channel of energy transport at CNP.  The inclusion of disorder, even in the smallest amount, ``cures" the singularities and creates a region of ``disorder-enabled hydrodynamics" in the immediate vicinity of the CNP.  Thus, the second purpose of this paper is to determine the qualitative behavior of the thermo-electric transport coefficients of bilayer graphene in this regime.  We find that the WF ratio, plotted as a function of doping density, follows a squared Lorentzian behavior, whose  quarter-maximum  occurs at doping density $n_c$ proportional to the strength of disorder, such that the WF ratio increases with decreasing disorder for $n<n_c$ and decreases with decreasing disorder for $n>n_c$.  At the same time, the Seebeck coefficient exhibits an interesting non-monotonic behavior, vanishishing at CNP and peaking in absolute value at $n\sim n_c$.  This behavior, admittedly very similar to what has been predicted and observed in single-layer graphene,  should be promptly comparable with the results of experimental measurements, as soon as they become available.  

Within the framework of quasi-classical transport theory \cite{ashcroft}, the state of the carriers is described by a non-equilibrium distribution function $f_{\kv,\gamma}$, where $\kv$ is the Bloch wave vector and $\gamma=\pm 1$ is the band index.  The deviation from equilibrium is $\delta f_{\kv,\gamma}= f_{\kv,\gamma}-f_{0\kv,\gamma}$, where $f_{0\kv,\gamma}$ is the equilibrium distribution function at chemical potential $\mu$ and temperature $T$.   The quantities of interest are the electric current $\jv_e$ and the thermal current $\jv_q$, however, in order to homogenize the dimensions we will be working with the particle current $\jv_n = \jv_e/(-e)$ ($-e$ is the charge of the electron) and the entropy current in units of the Boltzmann constant $k_B$,  $\jv_s  =\beta \jv_q$, where $\beta=(k_BT)^{-1}$. With this choice, the thermoelectric matrix [see below Eq.~(\ref{ResistivityMatrix})] manifestly satisfies Onsager reciprocity. The currents are related to the non-equilibrium distribution function by \cite{pines}
\be\label{EqJ}
\jv_n=\sum_{\kv,\gamma}\vv_{\kv,\gamma}\delta f_{\kv,\gamma},~~\jv_s=\sum_{\kv,\gamma}\beta\tilde{\epsilon}_{\kv,\gamma}\vv_{\kv,\gamma} \delta f_{\kv,\gamma}\,,
\ee
where $\epsilon_{\kv,\gamma}=\gamma[\sqrt{(t/2)^2+(\hbar v k)^2}-t/2]$ and  $\vv_{\kv,\gamma}=2\gamma\hbar v^2\kv/\sqrt{t^2+(2\hbar vk)^2}$ are, respectively, the energy and the velocity of band $\gamma = \pm$ at the wave vector $\kv$, while $\tilde\epsilon_{\kv,\gamma}\equiv\epsilon_{\kv,\gamma}-\mu$. Here $v = 10^8~{\rm cm/s}$ is the Fermi velocity and $t=0.4$ eV is the vertical interlayer hopping between the two layers~\cite{mccann_ssc_2007,mccann_rpp_2013}. Our two-band approximation is justified up to $T \sim t/k_{\rm B} \sim 4,600~{\rm K}$.

The currents are connected to the electric field $\Ev$ and to the temperature gradient $\nablabold T$ by the thermoelectric resistivity matrix, $\rhov$, which we define as
$ -\left(e\Ev~,~k_B\nablabold T\right)^{\mathrm{T}}=\rhov\cdot\left(\jv_n~,~ \jv_s\right)^{\mathrm{T}}$ (here ``T'' stands for the vector transposition).
 The elements of $\rhov$ are expressed in terms of three transport coefficients: the reduced electric resistivity $\bar \rho_{el}$, i.e. the ordinary electric resistivity multiplied by $e^2$, the reduced thermal resistivity $\bar \rho_{th}$, i.e. the usual thermal resistivity multiplied by $k_B^2 T$, and  the dimensionless Seebeck coefficient $\bar Q$, i.e. the ordinary Seebeck coefficient expressed in units of $k_B/e$, in the following form (see Ref.~\onlinecite{ashcroft})
\be\label{ResistivityMatrix}
\rhov=\left(\begin{array}{cc}\bar \rho_{el}+\bar Q^2\bar\rho_{th}& \bar Q \bar \rho_{th}\\ \bar Q \bar \rho_{th} & \bar\rho_{th}\end{array}\right)\,,
\ee
with ${\rm det} \rhov = \bar\rho_{el} \bar\rho_{th}$. The (dimensionless) Wiedemann-Franz ratio is defined as
$
WF \equiv
\bar \rho_{el}/\bar\rho_{th} =
{\rm det} \rhov/\bar\rho_{th}^2,
$.
$WF=\pi^2/3$ when the Wiedemann-Franz law is satisfied in its standard form, e.g., for a parabolic band electron gas in the presence of quenched short-range disorder. 
\begin{figure}[t]
\includegraphics[width=9.5cm]{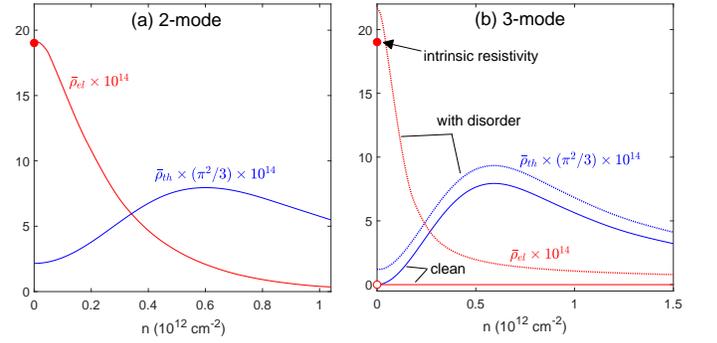}
\caption{
A comparison between the dimensionless electric and thermal resistivities $\bar \rho_{el}$ and $\bar \rho_{th}$, calculated with a (a) two-mode {\it Ansatz} and (b) three-mode {\it Ansatz}. Curves are plotted as functions of density (in units of $10^{12}~{\rm cm}^{-2}$ and for $T=100$ K. Solid curves correspond to the clean limit, whereas dotted ones are the calculated in the presence of small amount of disorder ($\lambda=0.005$).}
\label{fig1}
\end{figure}

\textit{Two-mode Ansatz.} The standard textbook calculation of thermoelectric coefficients~\cite{ashcroft} starts with the introduction of a 2-parameter {\it Ansatz} 
$
\delta f_{\kv,\gamma}=f^{\prime}_{0\kv,\gamma}\vv_{\kv,\gamma}\cdot (\pv_n+\beta\tilde\epsilon_{\kv,\gamma}\pv_s)
$,
where $\vv_{\kv,\gamma}$ and $\beta\tilde\epsilon_{\kv,\gamma}\vv_{\kv,\gamma}$ are the ``modes'' used to expand $\delta f_{\kv,\gamma}$. The two parameters $\pv_n$ and $\pv_s$ correspond to shifts of the particle momentum associated with the $\jv_n$ and $\jv_s$, respectively. Finally, the factor $f^{\prime}_{0\kv,\gamma}$, which denotes the derivative of the Fermi distribution with respect to energy, accounts for the fact that only electrons around the Fermi surface are mobile.

We stress that the choice of the {\it Ansatz} is the delicate point of the entire calculation. (The rest of the section is completely general and valid independently of such choice.)
The two-parameter {\it Ansatz} above is, for BLG, {\it incomplete} and leads to the {\it wrong} results for $\bar \rho_{el}$ and $\bar \rho_{th}$. In the following section we will amend it by including a third parameter, corresponding to the current conserved by the collision integral ({\it i.e.} the momentum density). 

Using Eq.~(\ref{EqJ}) we obtain $\left(\jv_n~,~\jv_s\right)^{\mathrm{T}}=\Dv \cdot\left(\pv_n~,~\pv_s\right)^{\mathrm{T}}$,
where $\Dv$ is the $2\times 2$ matrix of ``Drude weights"
\be \label{eq:Drude_weight}
D_{ij} = \frac{1}{2}\sum_{\kv,\gamma}f_{0\kv\gamma}'  \uv^i_{\kv,\gamma}\cdot \uv^j_{\kv,\gamma}
~,
\ee
where $\uv^i_{\kv,\gamma} = (\vv_{\kv,\gamma},\beta\tilde\epsilon_{\kv,\gamma}\vv_{\kv,\gamma})_i$ and $i,j=1,2$. $D_{ij}$ quantifies the ``{\it overlap}'' between the modes $\uv^i$ and $\uv^j$ (it can in fact be interpreted as a scalar product in the mode space).
To determine $\pv_n$ and $\pv_s$, we substitute the {\it Ansatz} for $\delta f_{\kv,\gamma}$ into the Boltzmann equation for the steady state response in the presence of fields $\Ev$ and $\nablabold T$,
\be \label{eq:Boltzmann_eq}
-f^{\prime}_{0\kv,\gamma}\vv_{\kv,\gamma}\cdot \left[e\Ev+\beta\tilde\epsilon_{\kv,\gamma}k_B\nablabold T\right]=I_{\kv,\gamma}
~.
\ee
Here $I_{\kv,\gamma}$ is the collision integral, which depends on the details of the microscopic scattering mechanism. 
Eq.~(\ref{eq:Boltzmann_eq}) is projected over the same set of modes which are used to expand $\delta f_{\kv,\gamma}$, {\it i.e.} it is multiplied by one of the modes $\uv^i_{\kv,\gamma}$, integrated over $\kv$ and summed over bands. In this way, the differential Eq.~(\ref{eq:Boltzmann_eq}) is transformed into an algebraic one, and is easily solved for $\pv_n$ and $\pv_s$.
The key inputs are the moments of the collision integral, which to linear order in $\pv_n$ and $\pv_s$ are given by
%
$\sum_{\kv,\gamma}\uv^i_{\kv,\gamma}I_{\kv,\gamma} = \tilde I_{ij} (\pv_n~,~\pv_s)_j$.
%
Hereafter summation of repeated latin indices is understood.
Such equation
defines the $2\times 2$ matrix $\tilde \Iv$, which we refer to as ``collision kernel", and whose matrix elements are the ${\tilde I}_{ij}$ ($i,j = 1,2$).
In the supplementary online material \cite{supp} we make use of a standard approximation for the Coulomb collision integral (screened interaction plus Fermi golden rule) to find
\be\label{ICC}
\tilde I_{ij}=-\frac{\beta }{4\pi }\sum_{\qv  }\int_{-\infty}^{\infty}d\omega\frac{|V(q)|^2 (\Im\Pi_{i}^{1} \Im\Pi_{j}^1-\Im \Pi^0\Im \Pi^2_{ij})}{\sinh^2(\beta\hbar\omega/2)},
\ee
where the response functions $\Pi^\alpha(\qv,\omega)$ are defined as
\be\label{EqPn}
\Pi^\alpha =4\sum_{\gamma,\gamma'}\sum_{\kv } \frac{F_{\kv ,\kv +\qv  }^{\gamma\gamma'}(f_{0\kv ,\gamma} -f_{0\kv +\qv  ,\gamma'})}{\epsilon_{\kv ,\gamma} -\epsilon_{\kv +\qv  ,\gamma'}+\hbar\omega+i0^{+}} M^\alpha_{\kv,\kv+\qv,\gamma,\gamma'}
.
\ee
Here $M^0_{\kv,\kv+\qv,\gamma,\gamma'} = 1$, $M^1_{\kv,\kv+\qv,\gamma,\gamma'} = \xv\cdot (\uv^i_{\kv ,\gamma} -\uv^i_{\kv +\qv  ,\gamma'})$, and $M^2_{\kv,\kv+\qv,\gamma,\gamma'} = (\uv^i_{\kv ,\gamma} -\uv^i_{\kv +\qv  ,\gamma'})\cdot(\uv^j_{\kv,\gamma} -\uv^j_{\kv+\qv,\gamma'})$.
Furthermore, $V(q)=2\pi e^2/[\kappa(q+q_{\mathrm{TF}})]$ is the screened Coulomb interaction, $q_{\mathrm{TF}}=4e^2\varepsilon_F/(\kappa\hbar^2 v^2 )$  is the Thomas-Fermi screening wave vector ($\varepsilon_F$ is the Fermi energy). The factor 4 accounts for the spin and valley degeneracy and $\kappa$ is the dielectric constant
of the substrate (within our calculations, we set $\kappa=4$ as for an h-BN substrate). 
The form factors $F_{\kv ,\kv +\qv}^{\gamma\gamma'}=[1+\gamma\gamma'\cos(2\theta_{\kv }-2\theta_{\kv +\qv  })]/2$, where $\theta_\kv$ is the angle formed by the $\kv$ vector with the $x$-axis, come from the overlap of the wave functions at wave vectors $\kv$ and $\kv+\qv$.  
%
%
\begin{figure*}
\includegraphics[width=14cm]{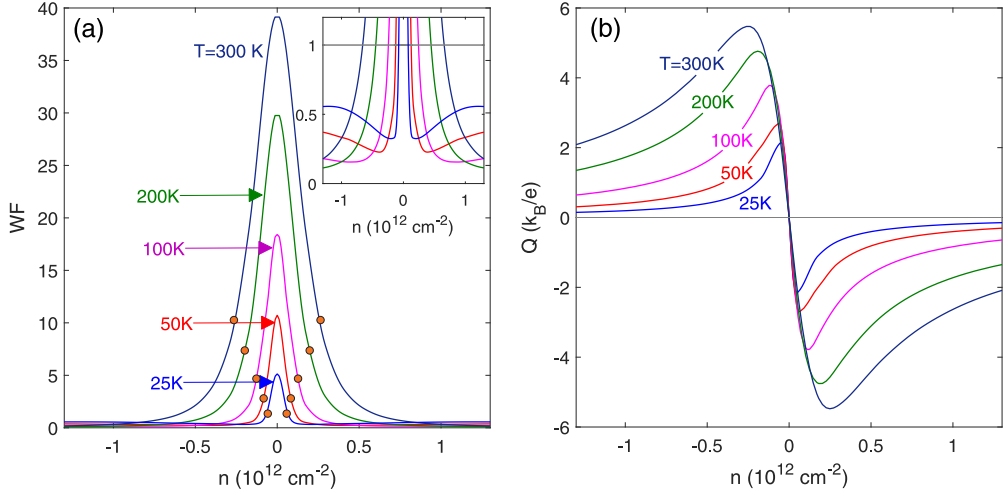}
\caption{ (a) Numerical WF ratio (scaled with $\pi^2/3$) and (b) Seebeck coefficient as  a function of density for different temperatures as labeled. The disorder strength is $\lambda=0.005$. The full red dots indicate the critical densities $n_c$ which are obtained by setting $\tilde\rho_D=\bar\rho_{el,C}$.}\label{fig2}
\end{figure*}
%
Simple algebraic manipulations lead to the final expression for the {\it two-mode} thermoelectric resistivity matrix,
$\tilde \rhov = \Dv^{-1}\cdot\tilde \Iv\cdot\Dv^{-1}$. For future purposes we define the conductivity matrix $\tilde \sigmabold \equiv \tilde \rhov^{-1} = \Dv\cdot\tilde \Iv^{-1}\cdot\Dv$.

This procedure works remarkably well for both parabolic-band electron gases and monolayer graphene in the clean limit.~\cite{zarenia} 
In such limit, {\it i.e.} when the only collision mechanism is the electron-electron interaction, the thermoelectric transport is strongly influenced by the exact conservation of the {\it momentum density}. 
For example, when one of the two currents overlaps sufficiently with the momentum density [where the overlap is defined as in Eq.~(\ref{eq:Drude_weight})], then it is also conserved. This in turn implies that either $\bar \rho_{el}$ or $\bar \rho_{th}$ vanishes. 
For parabolic-band electron gases and (massive or massless) Dirac systems, such result is readily obtained with a two-mode {\it Ansatz}. In such special cases, the overlap between the two currents and momentum (which coincides, in fact, with one of them at all times) is {\it automatically} taken into account.

It is then clear what fails when the two-mode {\it Ansatz} is applied to BLG: the overlap between the currents and the momentum density, the conserved mode of the collision integral, is not explicitly taken into account. Hence, all the coefficients of the matrix $\tilde \rhov$ are non-zero, and so is its determinant. Therefore, all thermoelectric transport coefficients are finite, see Fig.~\ref{fig1}(a). Such result is {\it wrong} as we proceed to show in the next section.


\textit{Three-mode Ansatz.} 
To account for the conserved mode of the collision integral (the momentum density)
we express the deviation of $f_{\kv,\gamma}$ from equilibrium as
\be\label{Ansatz3}
\delta f_{\kv,\gamma}=f^{\prime}_{0\kv,\gamma} \left[\hbar v^2\kv\cdot\pv_k/t+ \vv_{\kv,\gamma}\cdot(\pv_n+\beta\tilde\epsilon_{\kv,\gamma}\pv_s)\right]\,,
\ee
where $\pv_k$ is the shift associated with the momentum density scaled with $t/\hbar v^2$, in such a way that it has the same dimension as the other modes. 

A more general situation (away from the CNP) might require to include more modes, in order to achieve a better quantitative agreement with the true solution of the Boltzmann equation. However,  as we will shown later,  the {\it Ansatz} (\ref{Ansatz3}) is able to capture all  the main {\it qualitative} features (vanishing of the thermal/charge conductivities, divergence of the Seebeck coefficient, etc.). In addition, there is no real locking imposed {\it a priori} between the electron and hole velocities. As such, the three modes of the anstaz describe different situations, in which particles and holes may co-propagate, counter-propagate, or any combination in between. The coefficients of the separate modes are allowed to vanish, describing situations in which one particular configuration is realised.

We now define the three-component mode vector as $\uv^j_{\kv,\gamma} = (\hbar v^2\kv/t, \vv_{\kv,\gamma},\beta\tilde\epsilon_{\kv,\gamma}\vv_{\kv,\gamma})_j$. Using Eq.~(\ref{EqJ}) and~(\ref{eq:Drude_weight}), we get
$
(\jv_n, \jv_s)_i= D_{ij} (\pv_k, \pv_n, \pv_s)_j
$,
where $i=1,2$ and $j=0,1,2$,. Note that $\Dv$ is now a $2\times 3$ matrix.
Similarly, for the Coulomb collision kernel we find
\be
I_{jj'} = \left(\begin{array}{ccc} \eta &0&0\\0&\tilde I_{11}&\tilde I_{12}\\0&\tilde I_{12}&\tilde I_{22}\end{array}\right)_{jj'}
\ee
where $j,j' =0,1,2$. The $\tilde{I}_{ij}$ are defined in Eq.~(\ref{ICC}).
The conductivity matrix is easily seen to be 
$\sigmabold = \Dv\cdot \Iv^{-1}\cdot \Dv$.
%
Inverting it, in the limit of  $\eta\to0$, we obtain
 \be\label{RhoC3}
{\rhov} =
\frac{\left(\begin{array}{cc}
[D_{02}]^2 &-D_{01}D_{02}\\
-D_{01}D_{02}& [D_{01}]^2 
\end{array}\right)}
{[D_{01}]^2\tilde\sigma_{22}+[D_{02}]^2\tilde\sigma_{11}-2D_{01}D_{02} [\tilde\sigma_{12}]^2}
\ee
where $\tilde \sigma_{ii'} = \lim_{\eta\to 0} \sigma_{ii'}$ coincide with the conductivities obtained with the two-mode {\it Ansatz}.
Note that, since ${\rm det}\rhov=\bar\rho_{el}\bar\rho_{th}=0$,  either the electrical or thermal resistivity must necessarily be zero.
At charge neutrality $D_{01}=0$. Therefore,
${\bf \rho}_{11}(n=0) = \tilde\sigma_{11}^{-1}$, while all the other components of $\rhov$ are zero.
Thus, at the CNP, the thermal resistivity vanishes and the electrical resistivity is finite.
Away from the CNP we find
\be
\bar\rho_{th}(n\neq0)= \rho_{22},~\bar\rho_{el}(n\neq0)=0,~\bar Q=-D_{02}/D_{01}
~.
\ee

Numerical results for $\bar\rho_{el}$ and $\bar\rho_{th}$ are shown in Fig.~\ref{fig1}.
Within the two-mode approximation [Fig.~\ref{fig1}(a)] both the electrical and thermal resistivities are always finite.
In contrast, with the three mode {\it Ansatz} [Fig.~\ref{fig1}(b)], the electrical resistivity $\bar\rho_{el}(n\neq0)=0$ and exhibits a discontinuity at the CNP  in the absence of disorder (solid curves). 
Interestingly, both models yield the same intrinsic resistivity at $n=0$ [full red dots in Fig.~\ref{fig1}(a) and~(b)].  At  this point the only carriers in the system are thermally excited electrons and holes, in the conduction and valence bands, respectively. The two types of carries drift in opposite directions under the action of an electric field. Because of the transfer of momentum (known as Coulomb drag~\cite{milrinDrag}) between them, the resistivity becomes finite.

\textit{Disorder-enabled hydrodynamics.} 
An infinitesimal amount of disorder, which breaks the exact conservation of momentum, regularizes the singular results of the clean limit.  
We assume a momentum-non-conserving kernel $\Iv_D$ (D for disorder) proportional to a dimensionless momentum relaxation rate $\lambda$, which we take to be $\ll 1$.  The precise form of $\Iv_D$ is not important for our purposes. However, for the sake of  illustration, we will later make use of a simple model of electrons and holes scattering against randomly distributed impurities of density $n_d$ with short-range potential $V_0\delta(\rv)$: for this model $\lambda = n_dV_0^2/(\hbar v)^2$, as detailed in the supplementary online material \cite{supp}.  

The resistivity matrix is now the sum of two term $\rhov = \rhov_C + \rhov_D$, 
where $\rhov_C$ is given by Eq.~(\ref{RhoC3}), whereas $\rhov_D =  \Dv^{-1}\cdot  \Iv_D\cdot \Dv^{-1}$. 
The electrical and thermal resistivities now read~\cite{zarenia} 
$\bar\rho_{el}^{-1} \simeq \bar \rho_{el,C}^{-1}+\tilde\rho_{D}^{-1}$
and
$\bar\rho_{th} =\bar\rho_{th,C} +\bar\rho_{th,D}$,
where $\tilde\rho_{D} = \bar \rho_{th,D}(D_{02}/D_{01})^2$. In Fig.~\ref{fig1}(b) we plot such $\bar\rho_{el}$ and $\bar\rho_{th}$ (dotted curves) and compare them with the results of the previous section (solid curves). The effect of disorder on the thermal resistivity is just a small shift. Conversely, it regularizes the electric resistivity.
From the results above we obtain the WF ratio
\be\label{LorentzianSquared}
WF =\left(\frac{\Gamma}{(D_{01}/D_{02})^2+\Gamma^2}\right)^2\,,~~\Gamma^2=\frac{\bar \rho_{th,D}}{\bar\rho_{el,C}}\,.
\ee
which is a {\it square} of a Lorentzian. 
This formula shows that, at the CNP, $WF\to 1/\Gamma^2$, {\it i.e.} it is greatly enhanced relative to its standard noninteracting value $\pi^2/3$ and diverges as the strength of disorder tends to zero. 

In Figs.~\ref{fig2}(a) and~(b) we show numerical results for $WF$ and the Seebeck coefficient $\bar Q$, respectively.
We define a crossover density $n_c$, below which the enhancement of the WF persists, from the condition $\tilde\rho_D = \bar\rho_{el,C}$ [the dots in Fig.~\ref{fig2}(a) indicate its position]. 
We call such regime ($n<n_c$) ``{\it disorder-enabled hydrodynamics}''~\cite{zarenia}. In it, $WF$ remains much larger than $\pi^2/3$, in fact larger than $1/\Gamma^2 \gg \pi^2/3$.
(A more lenient crossover density could be defined~\cite{zarenia} as that at which $WF$ first drops below $\pi^2/3$.)

The disorder-regularized Seebeck coefficient [Figs.~\ref{fig2}(b)] exhibits a large swing about the CNP and goes to zero at $n=0$, as expected from particle-hole symmetry. The swing region, in which the derivative of $\bar Q$ vs density reverses its sign, is yet another incarnation of the disorder-enabled hydrodynamic regime~\cite{zarenia}. Its width is defined by the same condition $|n|<n_c$.

In Summary, we have presented the theory of thermoelectric transport in clean bilayer graphene. 
We have shown that the conventional semi-classical (Boltzmann) textbook approach to the calculation of thermoelectric coefficients~\cite{ashcroft}, which works remarkably well for both parabolic-band electron gases and Dirac systems~\cite{zarenia}, fails for such system. 
This is attributed to the fact that neither the particle current nor the energy current are conserved quantities in bilayer graphene. 
The correct results are found by explicitly including the current conserved by the electron-electron collision integral, {\it i.e.} the momentum density, in the ``augmented'' {\it Ansatz} for the non-equilibrium distribution function. Note that in previously studied examples the explicit addition of the conserved momentum mode was not required. This fortunate situation occurred because the
momentum mode was automatically subsumed under either the particle current or the energy current mode. Bilayer graphene is the first system studied in this context in which the two-mode Ansatz fails and the momentum mode must be introduced explicitly.

We find that, at the charge neutrality point: (i) the thermal resistivity vanishes; (ii) the electric resistivity jumps from zero to a finite value; (iii) the Seebeck coefficient diverges. These singularities are cured by the inclusion of a small amount of disorder. Breaking the exact momentum conservation, it opens a ``window" of doping densities $0<n<n_c$ (with $n_c$ tending to zero in the zero-disorder limit) in which the Wiedemann-Franz law is largely violated. There, the WF ratio greatly exceeds the standard value and the Seebeck coefficient exhibits a non-monotonic behavior (as a function of doping density). Such predictions can be tested in experiments in sufficiently clean samples of bilayer graphene.

\begin{acknowledgements}
{\it Acknowledgment.} This work was supported by the U.S. Department of Energy (Office of Science) under grant
No. DE-FG02-05ER46203. A.P. and T.B.S. acknowledge support from the Royal Society International Exchange grant IES\textbackslash R3\textbackslash 170252.
\end{acknowledgements}


\begin{thebibliography}{26}%
\makeatletter
\providecommand \@ifxundefined [1]{%
 \@ifx{#1\undefined}
}%
\providecommand \@ifnum [1]{%
 \ifnum #1\expandafter \@firstoftwo
 \else \expandafter \@secondoftwo
 \fi
}%
\providecommand \@ifx [1]{%
 \ifx #1\expandafter \@firstoftwo
 \else \expandafter \@secondoftwo
 \fi
}%
\providecommand \natexlab [1]{#1}%
\providecommand \enquote  [1]{``#1''}%
\providecommand \bibnamefont  [1]{#1}%
\providecommand \bibfnamefont [1]{#1}%
\providecommand \citenamefont [1]{#1}%
\providecommand \href@noop [0]{\@secondoftwo}%
\providecommand \href [0]{\begingroup \@sanitize@url \@href}%
\providecommand \@href[1]{\@@startlink{#1}\@@href}%
\providecommand \@@href[1]{\endgroup#1\@@endlink}%
\providecommand \@sanitize@url [0]{\catcode `\\12\catcode `\$12\catcode
  `\&12\catcode `\#12\catcode `\^12\catcode `\_12\catcode `\%12\relax}%
\providecommand \@@startlink[1]{}%
\providecommand \@@endlink[0]{}%
\providecommand \url  [0]{\begingroup\@sanitize@url \@url }%
\providecommand \@url [1]{\endgroup\@href {#1}{\urlprefix }}%
\providecommand \urlprefix  [0]{URL }%
\providecommand \Eprint [0]{\href }%
\providecommand \doibase [0]{http://dx.doi.org/}%
\providecommand \selectlanguage [0]{\@gobble}%
\providecommand \bibinfo  [0]{\@secondoftwo}%
\providecommand \bibfield  [0]{\@secondoftwo}%
\providecommand \translation [1]{[#1]}%
\providecommand \BibitemOpen [0]{}%
\providecommand \bibitemStop [0]{}%
\providecommand \bibitemNoStop [0]{.\EOS\space}%
\providecommand \EOS [0]{\spacefactor3000\relax}%
\providecommand \BibitemShut  [1]{\csname bibitem#1\endcsname}%
\let\auto@bib@innerbib\@empty
\bibitem [{\citenamefont {Narozhny}\ \emph {et~al.}(2017)\citenamefont
  {Narozhny}, \citenamefont {Gornyi}, \citenamefont {Mirlin},\ and\
  \citenamefont {Schmalian}}]{narozhnyRev}%
  \BibitemOpen
  \bibfield  {author} {\bibinfo {author} {\bibfnamefont {B.~N.}\ \bibnamefont
  {Narozhny}}, \bibinfo {author} {\bibfnamefont {I.~V.}\ \bibnamefont
  {Gornyi}}, \bibinfo {author} {\bibfnamefont {A.~D.}\ \bibnamefont {Mirlin}},
  \ and\ \bibinfo {author} {\bibfnamefont {J.}~\bibnamefont {Schmalian}},\
  }\href {\doibase 10.1002/andp.201700043} {\bibfield  {journal} {\bibinfo
  {journal} {Annalen der Physik}\ }\textbf {\bibinfo {volume} {529}},\ \bibinfo
  {pages} {1700043} (\bibinfo {year} {2017})}\BibitemShut {NoStop}%
\bibitem [{\citenamefont {Principi}\ \emph {et~al.}(2016)\citenamefont
  {Principi}, \citenamefont {Vignale}, \citenamefont {Carrega},\ and\
  \citenamefont {Polini}}]{Principi_prb_2016}%
  \BibitemOpen
  \bibfield  {author} {\bibinfo {author} {\bibfnamefont {A.}~\bibnamefont
  {Principi}}, \bibinfo {author} {\bibfnamefont {G.}~\bibnamefont {Vignale}},
  \bibinfo {author} {\bibfnamefont {M.}~\bibnamefont {Carrega}}, \ and\
  \bibinfo {author} {\bibfnamefont {M.}~\bibnamefont {Polini}},\ }\href
  {\doibase 10.1103/PhysRevB.93.125410} {\bibfield  {journal} {\bibinfo
  {journal} {Phys. Rev. B}\ }\textbf {\bibinfo {volume} {93}},\ \bibinfo
  {pages} {125410} (\bibinfo {year} {2016})}\BibitemShut {NoStop}%
\bibitem [{\citenamefont {Narozhny}\ \emph {et~al.}(2015)\citenamefont
  {Narozhny}, \citenamefont {Gornyi}, \citenamefont {Titov}, \citenamefont
  {Sch\"utt},\ and\ \citenamefont {Mirlin}}]{Narozhny_prb_2015}%
  \BibitemOpen
  \bibfield  {author} {\bibinfo {author} {\bibfnamefont {B.~N.}\ \bibnamefont
  {Narozhny}}, \bibinfo {author} {\bibfnamefont {I.~V.}\ \bibnamefont
  {Gornyi}}, \bibinfo {author} {\bibfnamefont {M.}~\bibnamefont {Titov}},
  \bibinfo {author} {\bibfnamefont {M.}~\bibnamefont {Sch\"utt}}, \ and\
  \bibinfo {author} {\bibfnamefont {A.~D.}\ \bibnamefont {Mirlin}},\ }\href
  {\doibase 10.1103/PhysRevB.91.035414} {\bibfield  {journal} {\bibinfo
  {journal} {Phys. Rev. B}\ }\textbf {\bibinfo {volume} {91}},\ \bibinfo
  {pages} {035414} (\bibinfo {year} {2015})}\BibitemShut {NoStop}%
\bibitem [{\citenamefont {Briskot}\ \emph {et~al.}(2015)\citenamefont
  {Briskot}, \citenamefont {Sch\"utt}, \citenamefont {Gornyi}, \citenamefont
  {Titov}, \citenamefont {Narozhny},\ and\ \citenamefont
  {Mirlin}}]{Briskot_prb_2015}%
  \BibitemOpen
  \bibfield  {author} {\bibinfo {author} {\bibfnamefont {U.}~\bibnamefont
  {Briskot}}, \bibinfo {author} {\bibfnamefont {M.}~\bibnamefont {Sch\"utt}},
  \bibinfo {author} {\bibfnamefont {I.~V.}\ \bibnamefont {Gornyi}}, \bibinfo
  {author} {\bibfnamefont {M.}~\bibnamefont {Titov}}, \bibinfo {author}
  {\bibfnamefont {B.~N.}\ \bibnamefont {Narozhny}}, \ and\ \bibinfo {author}
  {\bibfnamefont {A.~D.}\ \bibnamefont {Mirlin}},\ }\href {\doibase
  10.1103/PhysRevB.92.115426} {\bibfield  {journal} {\bibinfo  {journal} {Phys.
  Rev. B}\ }\textbf {\bibinfo {volume} {92}},\ \bibinfo {pages} {115426}
  (\bibinfo {year} {2015})}\BibitemShut {NoStop}%
\bibitem [{\citenamefont {Lucas}\ and\ \citenamefont
  {Fong}(2018)}]{lucasReview}%
  \BibitemOpen
  \bibfield  {author} {\bibinfo {author} {\bibfnamefont {A.}~\bibnamefont
  {Lucas}}\ and\ \bibinfo {author} {\bibfnamefont {K.~C.}\ \bibnamefont
  {Fong}},\ }\href {http://stacks.iop.org/0953-8984/30/i=5/a=053001} {\bibfield
   {journal} {\bibinfo  {journal} {Journal of Physics: Condensed Matter}\
  }\textbf {\bibinfo {volume} {30}},\ \bibinfo {pages} {053001} (\bibinfo
  {year} {2018})}\BibitemShut {NoStop}%
\bibitem [{\citenamefont {Bandurin}\ \emph {et~al.}(2016)\citenamefont
  {Bandurin}, \citenamefont {Torre}, \citenamefont {Kumar}, \citenamefont
  {Ben~Shalom}, \citenamefont {Tomadin}, \citenamefont {Principi},
  \citenamefont {Auton}, \citenamefont {Khestanova}, \citenamefont {Novoselov},
  \citenamefont {Grigorieva}, \citenamefont {Ponomarenko}, \citenamefont
  {Geim},\ and\ \citenamefont {Polini}}]{bandurin}%
  \BibitemOpen
  \bibfield  {author} {\bibinfo {author} {\bibfnamefont {D.~A.}\ \bibnamefont
  {Bandurin}}, \bibinfo {author} {\bibfnamefont {I.}~\bibnamefont {Torre}},
  \bibinfo {author} {\bibfnamefont {R.~K.}\ \bibnamefont {Kumar}}, \bibinfo
  {author} {\bibfnamefont {M.}~\bibnamefont {Ben~Shalom}}, \bibinfo {author}
  {\bibfnamefont {A.}~\bibnamefont {Tomadin}}, \bibinfo {author} {\bibfnamefont
  {A.}~\bibnamefont {Principi}}, \bibinfo {author} {\bibfnamefont {G.~H.}\
  \bibnamefont {Auton}}, \bibinfo {author} {\bibfnamefont {E.}~\bibnamefont
  {Khestanova}}, \bibinfo {author} {\bibfnamefont {K.~S.}\ \bibnamefont
  {Novoselov}}, \bibinfo {author} {\bibfnamefont {I.~V.}\ \bibnamefont
  {Grigorieva}}, \bibinfo {author} {\bibfnamefont {L.~A.}\ \bibnamefont
  {Ponomarenko}}, \bibinfo {author} {\bibfnamefont {A.~K.}\ \bibnamefont
  {Geim}}, \ and\ \bibinfo {author} {\bibfnamefont {M.}~\bibnamefont
  {Polini}},\ }\href {\doibase 10.1126/science.aad0201} {\bibfield  {journal}
  {\bibinfo  {journal} {Science}\ }\textbf {\bibinfo {volume} {351}},\ \bibinfo
  {pages} {1055} (\bibinfo {year} {2016})}\BibitemShut {NoStop}%
\bibitem [{\citenamefont {Ghahari}\ \emph {et~al.}(2016)\citenamefont
  {Ghahari}, \citenamefont {Xie}, \citenamefont {Taniguchi}, \citenamefont
  {Watanabe}, \citenamefont {Foster},\ and\ \citenamefont {Kim}}]{ghahari}%
  \BibitemOpen
  \bibfield  {author} {\bibinfo {author} {\bibfnamefont {F.}~\bibnamefont
  {Ghahari}}, \bibinfo {author} {\bibfnamefont {H.-Y.}\ \bibnamefont {Xie}},
  \bibinfo {author} {\bibfnamefont {T.}~\bibnamefont {Taniguchi}}, \bibinfo
  {author} {\bibfnamefont {K.}~\bibnamefont {Watanabe}}, \bibinfo {author}
  {\bibfnamefont {M.~S.}\ \bibnamefont {Foster}}, \ and\ \bibinfo {author}
  {\bibfnamefont {P.}~\bibnamefont {Kim}},\ }\href {\doibase
  10.1103/PhysRevLett.116.136802} {\bibfield  {journal} {\bibinfo  {journal}
  {Phys. Rev. Lett.}\ }\textbf {\bibinfo {volume} {116}},\ \bibinfo {pages}
  {136802} (\bibinfo {year} {2016})}\BibitemShut {NoStop}%
\bibitem [{\citenamefont {Krishna~Kumar}\ \emph {et~al.}(2017)\citenamefont
  {Krishna~Kumar}, \citenamefont {Bandurin}, \citenamefont {Pellegrino},
  \citenamefont {Cao}, \citenamefont {Principi}, \citenamefont {Guo},
  \citenamefont {Auton}, \citenamefont {Ben~Shalom}, \citenamefont
  {Ponomarenko}, \citenamefont {Falkovich}, \citenamefont {Watanabe},
  \citenamefont {Taniguchi}, \citenamefont {Grigorieva}, \citenamefont
  {Levitov}, \citenamefont {Polini},\ and\ \citenamefont {Geim}}]{kumar}%
  \BibitemOpen
  \bibfield  {author} {\bibinfo {author} {\bibfnamefont {R.}~\bibnamefont
  {Krishna~Kumar}}, \bibinfo {author} {\bibfnamefont {D.~A.}\ \bibnamefont
  {Bandurin}}, \bibinfo {author} {\bibfnamefont {F.~M.~D.}\ \bibnamefont
  {Pellegrino}}, \bibinfo {author} {\bibfnamefont {Y.}~\bibnamefont {Cao}},
  \bibinfo {author} {\bibfnamefont {A.}~\bibnamefont {Principi}}, \bibinfo
  {author} {\bibfnamefont {H.}~\bibnamefont {Guo}}, \bibinfo {author}
  {\bibfnamefont {G.~H.}\ \bibnamefont {Auton}}, \bibinfo {author}
  {\bibfnamefont {M.}~\bibnamefont {Ben~Shalom}}, \bibinfo {author}
  {\bibfnamefont {L.~A.}\ \bibnamefont {Ponomarenko}}, \bibinfo {author}
  {\bibfnamefont {G.}~\bibnamefont {Falkovich}}, \bibinfo {author}
  {\bibfnamefont {K.}~\bibnamefont {Watanabe}}, \bibinfo {author}
  {\bibfnamefont {T.}~\bibnamefont {Taniguchi}}, \bibinfo {author}
  {\bibfnamefont {I.~V.}\ \bibnamefont {Grigorieva}}, \bibinfo {author}
  {\bibfnamefont {L.~S.}\ \bibnamefont {Levitov}}, \bibinfo {author}
  {\bibfnamefont {M.}~\bibnamefont {Polini}}, \ and\ \bibinfo {author}
  {\bibfnamefont {A.~K.}\ \bibnamefont {Geim}},\ }\href
  {http://dx.doi.org/10.1038/nphys4240} {\bibfield  {journal} {\bibinfo
  {journal} {Nature Physics}\ }\textbf {\bibinfo {volume} {13}},\ \bibinfo
  {pages} {1182 } (\bibinfo {year} {2017})}\BibitemShut {NoStop}%
\bibitem [{\citenamefont {Gurzhi}(1963)}]{Gurzhi}%
  \BibitemOpen
  \bibfield  {author} {\bibinfo {author} {\bibfnamefont {R.}~\bibnamefont
  {Gurzhi}},\ }\href@noop {} {\bibfield  {journal} {\bibinfo  {journal} {Sov.
  Phys. JETP}\ }\textbf {\bibinfo {volume} {44}},\ \bibinfo {pages} {771}
  (\bibinfo {year} {1963})}\BibitemShut {NoStop}%
\bibitem [{\citenamefont {Zarenia}\ \emph {et~al.}(2018)\citenamefont
  {Zarenia}, \citenamefont {Principi},\ and\ \citenamefont
  {Vignale}}]{zarenia}%
  \BibitemOpen
  \bibfield  {author} {\bibinfo {author} {\bibfnamefont {M.}~\bibnamefont
  {Zarenia}}, \bibinfo {author} {\bibfnamefont {A.}~\bibnamefont {Principi}}, \
  and\ \bibinfo {author} {\bibfnamefont {G.}~\bibnamefont {Vignale}},\
  }\href@noop {} {\bibfield  {journal} {\bibinfo  {journal} {arXiv:1811.08914}\
  } (\bibinfo {year} {2018})}\BibitemShut {NoStop}%
\bibitem [{\citenamefont {Fritz}\ \emph {et~al.}(2008)\citenamefont {Fritz},
  \citenamefont {Schmalian}, \citenamefont {M\"uller},\ and\ \citenamefont
  {Sachdev}}]{Fritz_2008}%
  \BibitemOpen
  \bibfield  {author} {\bibinfo {author} {\bibfnamefont {L.}~\bibnamefont
  {Fritz}}, \bibinfo {author} {\bibfnamefont {J.}~\bibnamefont {Schmalian}},
  \bibinfo {author} {\bibfnamefont {M.}~\bibnamefont {M\"uller}}, \ and\
  \bibinfo {author} {\bibfnamefont {S.}~\bibnamefont {Sachdev}},\ }\href
  {\doibase 10.1103/PhysRevB.78.085416} {\bibfield  {journal} {\bibinfo
  {journal} {Phys. Rev. B}\ }\textbf {\bibinfo {volume} {78}},\ \bibinfo
  {pages} {085416} (\bibinfo {year} {2008})}\BibitemShut {NoStop}%
\bibitem [{\citenamefont {M\"uller}\ \emph {et~al.}(2008)\citenamefont
  {M\"uller}, \citenamefont {Fritz},\ and\ \citenamefont
  {Sachdev}}]{Muller_2008}%
  \BibitemOpen
  \bibfield  {author} {\bibinfo {author} {\bibfnamefont {M.}~\bibnamefont
  {M\"uller}}, \bibinfo {author} {\bibfnamefont {L.}~\bibnamefont {Fritz}}, \
  and\ \bibinfo {author} {\bibfnamefont {S.}~\bibnamefont {Sachdev}},\ }\href
  {\doibase 10.1103/PhysRevB.78.115406} {\bibfield  {journal} {\bibinfo
  {journal} {Phys. Rev. B}\ }\textbf {\bibinfo {volume} {78}},\ \bibinfo
  {pages} {115406} (\bibinfo {year} {2008})}\BibitemShut {NoStop}%
\bibitem [{\citenamefont {Svintsov}\ \emph {et~al.}(2012)\citenamefont
  {Svintsov}, \citenamefont {Vyurkov}, \citenamefont {Yurchenko}, \citenamefont
  {Otsuji},\ and\ \citenamefont {Ryzhii}}]{svintsov2012}%
  \BibitemOpen
  \bibfield  {author} {\bibinfo {author} {\bibfnamefont {D.}~\bibnamefont
  {Svintsov}}, \bibinfo {author} {\bibfnamefont {V.}~\bibnamefont {Vyurkov}},
  \bibinfo {author} {\bibfnamefont {S.}~\bibnamefont {Yurchenko}}, \bibinfo
  {author} {\bibfnamefont {T.}~\bibnamefont {Otsuji}}, \ and\ \bibinfo {author}
  {\bibfnamefont {V.}~\bibnamefont {Ryzhii}},\ }\href {\doibase
  10.1063/1.4705382} {\bibfield  {journal} {\bibinfo  {journal} {Journal of
  Applied Physics}\ }\textbf {\bibinfo {volume} {111}},\ \bibinfo {pages}
  {083715} (\bibinfo {year} {2012})}\BibitemShut {NoStop}%
\bibitem [{\citenamefont {Crossno}\ \emph {et~al.}(2016)\citenamefont
  {Crossno}, \citenamefont {Shi}, \citenamefont {Wang}, \citenamefont {Liu},
  \citenamefont {Harzheim}, \citenamefont {Lucas}, \citenamefont {Sachdev},
  \citenamefont {Kim}, \citenamefont {Taniguchi}, \citenamefont {Watanabe},
  \citenamefont {Ohki},\ and\ \citenamefont {Fong}}]{crossno}%
  \BibitemOpen
  \bibfield  {author} {\bibinfo {author} {\bibfnamefont {J.}~\bibnamefont
  {Crossno}}, \bibinfo {author} {\bibfnamefont {J.~K.}\ \bibnamefont {Shi}},
  \bibinfo {author} {\bibfnamefont {K.}~\bibnamefont {Wang}}, \bibinfo {author}
  {\bibfnamefont {X.}~\bibnamefont {Liu}}, \bibinfo {author} {\bibfnamefont
  {A.}~\bibnamefont {Harzheim}}, \bibinfo {author} {\bibfnamefont
  {A.}~\bibnamefont {Lucas}}, \bibinfo {author} {\bibfnamefont
  {S.}~\bibnamefont {Sachdev}}, \bibinfo {author} {\bibfnamefont
  {P.}~\bibnamefont {Kim}}, \bibinfo {author} {\bibfnamefont {T.}~\bibnamefont
  {Taniguchi}}, \bibinfo {author} {\bibfnamefont {K.}~\bibnamefont {Watanabe}},
  \bibinfo {author} {\bibfnamefont {T.~A.}\ \bibnamefont {Ohki}}, \ and\
  \bibinfo {author} {\bibfnamefont {K.~C.}\ \bibnamefont {Fong}},\ }\href
  {\doibase 10.1126/science.aad0343} {\bibfield  {journal} {\bibinfo  {journal}
  {Science}\ } (\bibinfo {year} {2016}),\ 10.1126/science.aad0343}\BibitemShut
  {NoStop}%
\bibitem [{\citenamefont {Principi}\ and\ \citenamefont
  {Vignale}(2015)}]{Principi_conductivity}%
  \BibitemOpen
  \bibfield  {author} {\bibinfo {author} {\bibfnamefont {A.}~\bibnamefont
  {Principi}}\ and\ \bibinfo {author} {\bibfnamefont {G.}~\bibnamefont
  {Vignale}},\ }\href {\doibase 10.1103/PhysRevB.91.205423} {\bibfield
  {journal} {\bibinfo  {journal} {Phys. Rev. B}\ }\textbf {\bibinfo {volume}
  {91}},\ \bibinfo {pages} {205423} (\bibinfo {year} {2015})}\BibitemShut
  {NoStop}%
\bibitem [{\citenamefont {Lucas}\ and\ \citenamefont
  {Das~Sarma}(2018)}]{lucasWF}%
  \BibitemOpen
  \bibfield  {author} {\bibinfo {author} {\bibfnamefont {A.}~\bibnamefont
  {Lucas}}\ and\ \bibinfo {author} {\bibfnamefont {S.}~\bibnamefont
  {Das~Sarma}},\ }\href {\doibase 10.1103/PhysRevB.97.245128} {\bibfield
  {journal} {\bibinfo  {journal} {Phys. Rev. B}\ }\textbf {\bibinfo {volume}
  {97}},\ \bibinfo {pages} {245128} (\bibinfo {year} {2018})}\BibitemShut
  {NoStop}%
\bibitem [{\citenamefont {Lucas}\ and\ \citenamefont
  {Hartnoll}(2018)}]{lucasKT}%
  \BibitemOpen
  \bibfield  {author} {\bibinfo {author} {\bibfnamefont {A.}~\bibnamefont
  {Lucas}}\ and\ \bibinfo {author} {\bibfnamefont {S.~A.}\ \bibnamefont
  {Hartnoll}},\ }\href {\doibase 10.1103/PhysRevB.97.045105} {\bibfield
  {journal} {\bibinfo  {journal} {Phys. Rev. B}\ }\textbf {\bibinfo {volume}
  {97}},\ \bibinfo {pages} {045105} (\bibinfo {year} {2018})}\BibitemShut
  {NoStop}%
\bibitem [{\citenamefont {Xie}\ and\ \citenamefont {Foster}(2016)}]{foster}%
  \BibitemOpen
  \bibfield  {author} {\bibinfo {author} {\bibfnamefont {H.-Y.}\ \bibnamefont
  {Xie}}\ and\ \bibinfo {author} {\bibfnamefont {M.~S.}\ \bibnamefont
  {Foster}},\ }\href {\doibase 10.1103/PhysRevB.93.195103} {\bibfield
  {journal} {\bibinfo  {journal} {Phys. Rev. B}\ }\textbf {\bibinfo {volume}
  {93}},\ \bibinfo {pages} {195103} (\bibinfo {year} {2016})}\BibitemShut
  {NoStop}%
\bibitem [{\citenamefont {McCann}\ and\ \citenamefont
  {Koshino}(2013{\natexlab{a}})}]{McCann}%
  \BibitemOpen
  \bibfield  {author} {\bibinfo {author} {\bibfnamefont {E.}~\bibnamefont
  {McCann}}\ and\ \bibinfo {author} {\bibfnamefont {M.}~\bibnamefont
  {Koshino}},\ }\href {http://stacks.iop.org/0034-4885/76/i=5/a=056503}
  {\bibfield  {journal} {\bibinfo  {journal} {Reports on Progress in Physics}\
  }\textbf {\bibinfo {volume} {76}},\ \bibinfo {pages} {056503} (\bibinfo
  {year} {2013}{\natexlab{a}})}\BibitemShut {NoStop}%
\bibitem [{\citenamefont {Zhang}\ \emph {et~al.}(2009)\citenamefont {Zhang},
  \citenamefont {Tang}, \citenamefont {Girit}, \citenamefont {Hao},
  \citenamefont {Martin}, \citenamefont {Zettl}, \citenamefont {Crommie},
  \citenamefont {Shen},\ and\ \citenamefont {Wang}}]{Zhang2009}%
  \BibitemOpen
  \bibfield  {author} {\bibinfo {author} {\bibfnamefont {Y.}~\bibnamefont
  {Zhang}}, \bibinfo {author} {\bibfnamefont {T.-T.}\ \bibnamefont {Tang}},
  \bibinfo {author} {\bibfnamefont {C.}~\bibnamefont {Girit}}, \bibinfo
  {author} {\bibfnamefont {Z.}~\bibnamefont {Hao}}, \bibinfo {author}
  {\bibfnamefont {M.~C.}\ \bibnamefont {Martin}}, \bibinfo {author}
  {\bibfnamefont {A.}~\bibnamefont {Zettl}}, \bibinfo {author} {\bibfnamefont
  {M.~F.}\ \bibnamefont {Crommie}}, \bibinfo {author} {\bibfnamefont {Y.~R.}\
  \bibnamefont {Shen}}, \ and\ \bibinfo {author} {\bibfnamefont
  {F.}~\bibnamefont {Wang}},\ }\href {https://doi.org/10.1038/nature08105}
  {\bibfield  {journal} {\bibinfo  {journal} {Nature}\ }\textbf {\bibinfo
  {volume} {459}},\ \bibinfo {pages} {820 EP } (\bibinfo {year}
  {2009})}\BibitemShut {NoStop}%
\bibitem [{\citenamefont {Ashcroft}\ and\ \citenamefont
  {Mermin}(2011)}]{ashcroft}%
  \BibitemOpen
  \bibfield  {author} {\bibinfo {author} {\bibfnamefont {N.}~\bibnamefont
  {Ashcroft}}\ and\ \bibinfo {author} {\bibfnamefont {N.}~\bibnamefont
  {Mermin}},\ }\href {https://books.google.com/books?id=x\_s\_YAAACAAJ} {\emph
  {\bibinfo {title} {Solid State Physics}}}\ (\bibinfo  {publisher} {Cengage
  Learning},\ \bibinfo {year} {2011})\BibitemShut {NoStop}%
\bibitem [{\citenamefont {Pines}\ and\ \citenamefont
  {Nozi{\`e}res}(1966)}]{pines}%
  \BibitemOpen
  \bibfield  {author} {\bibinfo {author} {\bibfnamefont {D.}~\bibnamefont
  {Pines}}\ and\ \bibinfo {author} {\bibfnamefont {P.}~\bibnamefont
  {Nozi{\`e}res}},\ }\href {https://books.google.com.sg/books?id=GP1QAAAAMAAJ}
  {\emph {\bibinfo {title} {The Theory of Quantum Liquids: Normal Fermi
  liquids}}}\ (\bibinfo  {publisher} {W.A. Benjamin},\ \bibinfo {year}
  {1966})\BibitemShut {NoStop}%
\bibitem [{\citenamefont {McCann}\ \emph {et~al.}(2007)\citenamefont {McCann},
  \citenamefont {Abergel},\ and\ \citenamefont {Fal’ko}}]{mccann_ssc_2007}%
  \BibitemOpen
  \bibfield  {author} {\bibinfo {author} {\bibfnamefont {E.}~\bibnamefont
  {McCann}}, \bibinfo {author} {\bibfnamefont {D.~S.}\ \bibnamefont {Abergel}},
  \ and\ \bibinfo {author} {\bibfnamefont {V.~I.}\ \bibnamefont {Fal’ko}},\
  }\href {\doibase https://doi.org/10.1016/j.ssc.2007.03.054} {\bibfield
  {journal} {\bibinfo  {journal} {Solid State Communications}\ }\textbf
  {\bibinfo {volume} {143}},\ \bibinfo {pages} {110 } (\bibinfo {year}
  {2007})}\BibitemShut {NoStop}%
\bibitem [{\citenamefont {McCann}\ and\ \citenamefont
  {Koshino}(2013{\natexlab{b}})}]{mccann_rpp_2013}%
  \BibitemOpen
  \bibfield  {author} {\bibinfo {author} {\bibfnamefont {E.}~\bibnamefont
  {McCann}}\ and\ \bibinfo {author} {\bibfnamefont {M.}~\bibnamefont
  {Koshino}},\ }\href {http://stacks.iop.org/0034-4885/76/i=5/a=056503}
  {\bibfield  {journal} {\bibinfo  {journal} {Reports on Progress in Physics}\
  }\textbf {\bibinfo {volume} {76}},\ \bibinfo {pages} {056503} (\bibinfo
  {year} {2013}{\natexlab{b}})}\BibitemShut {NoStop}%
\bibitem [{sup()}]{supp}%
  \BibitemOpen
  \href@noop {} {\ }\bibinfo {note} {See Supplemental Material at [URL will be
  inserted by publisher] for more details.}\BibitemShut {Stop}%
\bibitem [{\citenamefont {Sch\"utt}\ \emph {et~al.}(2013)\citenamefont
  {Sch\"utt}, \citenamefont {Ostrovsky}, \citenamefont {Titov}, \citenamefont
  {Gornyi}, \citenamefont {Narozhny},\ and\ \citenamefont
  {Mirlin}}]{milrinDrag}%
  \BibitemOpen
  \bibfield  {author} {\bibinfo {author} {\bibfnamefont {M.}~\bibnamefont
  {Sch\"utt}}, \bibinfo {author} {\bibfnamefont {P.~M.}\ \bibnamefont
  {Ostrovsky}}, \bibinfo {author} {\bibfnamefont {M.}~\bibnamefont {Titov}},
  \bibinfo {author} {\bibfnamefont {I.~V.}\ \bibnamefont {Gornyi}}, \bibinfo
  {author} {\bibfnamefont {B.~N.}\ \bibnamefont {Narozhny}}, \ and\ \bibinfo
  {author} {\bibfnamefont {A.~D.}\ \bibnamefont {Mirlin}},\ }\href {\doibase
  10.1103/PhysRevLett.110.026601} {\bibfield  {journal} {\bibinfo  {journal}
  {Phys. Rev. Lett.}\ }\textbf {\bibinfo {volume} {110}},\ \bibinfo {pages}
  {026601} (\bibinfo {year} {2013})}\BibitemShut {NoStop}%
\end{thebibliography}


%

\newpage

\begin{widetext}

\section*{Supplemental Material\\
Breakdown of the Wiedemann-Franz law in AB-stacked bilayer graphene}

In this supplemental material, we present more details on our  calculations for the electron-electron Coulomb (Sec. A) as well as our short-range model for the disorder collision moments (Sec. B).
\section*{A. Electron-electron collision moments}
\setcounter{equation}{0}
\renewcommand{\theequation}{A.\arabic{equation}}
The electron-electron collision integral for the band $\gamma$ and at wave vector $\kv$, $I_{\kv ,\gamma}$ is given by \cite{ashcroft,pines},
\ber
\tilde I_{\kv,\gamma} &=-\sum_{\kv'}\sum_{\gamma',\eta,\eta'}\sum_{\qv }W(q) 
\big[ f_{\kv,\gamma } (1-f_{\kv  -\qv,\gamma'})f_{\kv',\eta} (1-f_{\kv  '+\qv,\eta' })-f_{\kv -\qv,\gamma' }(1-f_{\kv,\gamma})f_{\kv '+\qv,\eta' }(1-f_{\kv',\eta} )\big]\times\nonumber\\
&\delta(\epsilon_{\kv,\gamma}  +\epsilon_{\kv',\eta} -\epsilon_{\kv  -\qv,\gamma' }-\epsilon_{\kv  +\qv ,\eta'})
\eer
where the momentum conservation appears naturally when doing the second quantization
of Coulomb interaction in $\kv$-space, and the energy conservation stems from the Fermi golden
rule. $f_{\kv,\gamma}=f_{0\kv,\gamma}+\delta f_{\kv,\gamma}$ is the non-equilibrium distribution function and $W(q)=(2\pi/\hbar)|V(q)|^2$ defines the collision probability where $V(\qv )$ is the statistic screened Coulomb interaction. Having the two-mode Ansatz
\be
\delta f_{\kv,\gamma}=f^{\prime}_{0\kv,\gamma}\vv_{\kv,\gamma}\cdot (\pv_n+\beta\tilde\epsilon_{\kv,\gamma}\pv_s)
\ee
and using
\be
\delta(\epsilon_{\kv,\gamma}  +\epsilon_{\kv',\eta} -\epsilon_{\kv  -\qv,\gamma' }-\epsilon_{\kv  +\qv ,\eta'})=\hbar\int_{-\infty}^{\infty}d\omega
\delta(\epsilon_{\kv',\eta} -\epsilon_{\kv  '+\qv,\eta' }+\hbar\omega)\delta(\epsilon_{\kv,\gamma } -\epsilon_{\kv  -\qv,\gamma' }-\hbar\omega)
\ee
and
\be
f_{0\kv,\gamma} (1-f_{0\kv\pm\qv,\gamma' })\delta(\epsilon_{\kv,\gamma} -\epsilon_{\kv\pm\qv,\gamma' }\pm\hbar\omega)=
\frac{f_{0\kv,\gamma} -f_{0\kv\pm\qv,\gamma' }}{\mp2e^{\pm\beta\hbar\omega/2}\sinh(\beta\hbar\omega/2)}\delta(\epsilon_{\kv,\gamma} -\epsilon_{\kv\pm\qv,\gamma' }\pm\hbar\omega),
\ee
we obtain
\ber\label{EqIk1}
&\tilde I_{\kv ,\gamma}= \frac{2\pi \beta}{4}\sum_{\gamma',\eta,\eta'}\sum_{\kv^{\prime} }\sum_{\qv  }\int_{-\infty}^{\infty}d\omega\frac{|V(\qv  )|^2}{\sinh^2(\beta\hbar\omega/2)}\times\nonumber\\
& F_{\kv ,\kv -\qv  }^{\gamma\gamma '}(f_{0\kv ,\gamma } -f_{0\kv -\qv  ,\gamma '})
\delta(\epsilon_{\kv ,\gamma } -\epsilon_{\kv -\qv  ,\gamma '}-\hbar\omega)
F_{\kv ',\kv '+\qv  }^{\eta\eta' }(f_{0\kv ',\eta } -f_{0\kv '+\qv  ,\eta '})\delta(\epsilon_{\kv ',\eta } -\epsilon_{\kv '+\qv  ,\eta '}+\hbar\omega)\times\nonumber\\
&\left[(\vv_{\kv ,\gamma }  -\vv_{\kv -\qv  ,\gamma '}+\vv_{\kv ',\eta } -\vv_{\kv '+\qv  ,\eta '})\right.\pv_n+
 \beta\left.(\vv_{\kv ,\gamma }\tilde{\epsilon}_{\kv ,\gamma }  -\vv_{\kv -\qv  ,\gamma '}\tilde{\epsilon}_{\kv -\qv  ,\gamma '}+\vv_{\kv ',\eta }\tilde{\epsilon}_{\kv ',\eta } -\vv_{\kv '+\qv  ,\eta '}\tilde{\epsilon}_{\kv '+\qv  ,\eta '})\right]\pv_s,
\eer
where $\{\gamma ,\gamma ',\eta ,\eta '\}=\pm 1$ denote the band index and $F_{\kv ,\kv \pm\qv  }^{\gamma\gamma '}$ is the form factor comes from the overlap of the wave functions at vector $\kv$ and $\kv\pm\qv$.  Now the moments of the electron-electron collision integrals are calculated using
\be\label{IMoments}
\left(\begin{array}{c}\sum_{\kv,\gamma}\vv_{\kv,\gamma}\tilde I_{\kv,\gamma} \\
\sum_{\kv,\gamma}\beta\tilde\epsilon_{\kv,\gamma}\vv_{\kv,\gamma}\tilde I_{\kv,\gamma}
\end{array}\right)=\tilde\Iv \cdot\left(
  \begin{array}{c}
    \pv_n\\
  \pv_s
  \end{array}
\right).
\ee
Using Eq. (\ref{EqIk1}) we write, 
\be\label{IMoments2}
\left(\begin{array}{c}\sum_{\kv,\gamma}\vv_{\kv,\gamma}\tilde I_{\kv,\gamma} \\
\sum_{\kv,\gamma}\beta\tilde\epsilon_{\kv,\gamma}\vv_{\kv,\gamma}\tilde I_{\kv,\gamma}
\end{array}\right)=\frac{\pi \beta}{2}\sum_{\qv  }\int_{-\infty}^{\infty}d\omega\frac{|V(\qv  )|^2}{\sinh^2(\beta\hbar\omega/2)}
\left(\begin{array}{cc}
G_{11}&G_{12}\\
G_{21}&G_{22}
\end{array}\right)
 \cdot\left(
  \begin{array}{c}
    \pv_n\\
  \pv_s
  \end{array}
\right),
\ee
where the matrix elements $G_{ij}(\qv,\omega,T)$ are
\ber\label{IG}
&G_{ij}= 
\left(\sum_{\kv,\gamma,\gamma'}u_{\kv,\gamma}^i F_{\kv ,\kv -\qv  }^{\gamma\gamma '}(f_{0\kv ,\gamma } -f_{0\kv -\qv  ,\gamma '})\delta(\epsilon_{\kv ,\gamma } -\epsilon_{\kv -\qv  ,\gamma '}-\hbar\omega)\right)\times\nonumber\\
&\left(\sum_{\kv',\eta,\eta'} (u_{\kv ',\eta }^j -u_{\kv '+\qv  ,\eta '}^j) F_{\kv ',\kv '+\qv  }^{\eta\eta' }(f_{0\kv ',\eta } -f_{0\kv '+\qv  ,\eta '})\delta(\epsilon_{\kv ',\eta } -\epsilon_{\kv '+\qv  ,\eta '}+\hbar\omega)\right)+ \nonumber\\
&\left(\sum_{\kv,\gamma,\gamma'}u_{\kv,\gamma}^i (u_{\kv ,\gamma }^j  -u_{\kv -\qv  ,\gamma'}^j)F_{\kv ,\kv -\qv  }^{\gamma\gamma '}(f_{0\kv ,\gamma } -f_{0\kv -\qv  ,\gamma '})\delta(\epsilon_{\kv ,\gamma } -\epsilon_{\kv -\qv  ,\gamma '}-\hbar\omega)\right)\times\nonumber\\
&\left(\sum_{\kv',\eta,\eta'}  F_{\kv ',\kv '+\qv  }^{\eta\eta' }(f_{0\kv ',\eta } -f_{0\kv '+\qv  ,\eta '})\delta(\epsilon_{\kv ',\eta } -\epsilon_{\kv '+\qv  ,\eta '}+\hbar\omega)\right).
\eer
As defined in the main text, the two-component vector $\uv^i_{\kv,\gamma} = (\vv_{\kv,\gamma},\beta\tilde\epsilon_{\kv,\gamma}\vv_{\kv,\gamma})_i$ ($i=1,2$) . By simple algebraic manipulations, one can show that the first and the third lines in Eq. (\ref{IG}) are equivalent to 
\ber\label{App1}
&\sum_{\kv,\gamma ,\gamma '}u_{\kv ,\gamma }^iF_{\kv ,\kv -\qv  }^{\gamma\gamma '} (f^0_{\kv ,\gamma } -f^0_{\kv -\qv  ,\gamma '})\delta(\epsilon_{\kv ,\gamma } -\epsilon_{\kv -\qv  ,\gamma '}-\hbar\omega)
=\nonumber\\
&\frac{1}{2}\sum_{\kv,\gamma ,\gamma '}F_{\kv ,\kv +\qv  }^{\gamma\gamma '}( u_{\kv ,\gamma }^i- u_{\kv +\qv  ,\gamma '}^i) (f^0_{\kv ,\gamma }-f^0_{\kv +\qv  ,\gamma '} )\delta(\epsilon_{\kv ,\gamma }-\epsilon_{\kv +\qv  ,\gamma '} +\hbar\omega)
\eer
and 
\ber\label{App2}
&\sum_{\kv,\gamma ,\gamma'}u_{\kv ,\gamma }^i(u_{\kv ,\gamma }^j -u_{\kv -\qv  ,\gamma '}^j)F_{\kv ,\kv -\qv  }^{\gamma\gamma '} (f^0_{\kv ,\gamma } -f^0_{\kv -\qv  ,\gamma '})\delta(\epsilon_{\kv ,\gamma } -\epsilon_{\kv -\qv  ,\gamma '}-\hbar\omega)=\nonumber\\
&-\frac{1}{2}\sum_{\kv,\gamma ,\gamma '} (u_{\kv ,\gamma }^i -u_{\kv +\qv  ,\gamma '}^i)(u_{\kv ,\gamma }^j -u_{\kv +\qv  ,\gamma '}^j)F_{\kv ,\kv +\qv  }^{\gamma\gamma '}(f^0_{\kv ,\gamma } -f^0_{\kv +\qv  ,\gamma '})\delta(\epsilon_{\kv ,\gamma } -\epsilon_{\kv +\qv  ,\gamma '}+\hbar\omega).
\eer
Inserting Eqs. (\ref{App1}) and (\ref{App2}) into Eq. (\ref{IG}) leads to
\ber\label{IG2}
&G_{ij}= 
\frac{1}{2}\left(\sum_{\kv,\gamma,\gamma'}(u_{\kv,\gamma}^i-u_{\kv+\qv,\gamma'}^i) F_{\kv ,\kv +\qv  }^{\gamma\gamma '}(f_{0\kv ,\gamma } -f_{0\kv +\qv  ,\gamma '})\delta(\epsilon_{\kv ,\gamma } -\epsilon_{\kv +\qv  ,\gamma '}+\hbar\omega)\right)\times\nonumber\\
&\left(\sum_{\kv',\eta,\eta'} (u_{\kv ',\eta }^j -u_{\kv '+\qv  ,\eta '}^j) F_{\kv ',\kv '+\qv  }^{\eta\eta' }(f_{0\kv ',\eta } -f_{0\kv '+\qv  ,\eta '})\delta(\epsilon_{\kv ',\eta } -\epsilon_{\kv '+\qv  ,\eta '}+\hbar\omega)\right)- \nonumber\\
&\frac{1}{2}\left(\sum_{\kv,\gamma ,\gamma '} (u_{\kv ,\gamma }^i -u_{\kv +\qv  ,\gamma '}^i)(u_{\kv ,\gamma }^j -u_{\kv +\qv  ,\gamma '}^j)F_{\kv ,\kv +\qv  }^{\gamma\gamma '}(f^0_{\kv ,\gamma } -f^0_{\kv +\qv  ,\gamma '})\delta(\epsilon_{\kv ,\gamma } -\epsilon_{\kv +\qv  ,\gamma '}+\hbar\omega)\right)\times\nonumber\\
&\left(\sum_{\kv',\eta,\eta'}  F_{\kv ',\kv '+\qv  }^{\eta\eta' }(f_{0\kv ',\eta } -f_{0\kv '+\qv  ,\eta '})\delta(\epsilon_{\kv ',\eta } -\epsilon_{\kv '+\qv  ,\eta '}+\hbar\omega)\right). 
\eer
Using this equation, $\tilde\Iv$ becomes, 
\be\label{ICC}
\tilde\Iv=-\frac{\beta }{4\pi }\sum_{\qv  }\int_{-\infty}^{\infty}d\omega\frac{|V(q)|^2}{\sinh^2(\beta\hbar\omega/2)}
\left(
\begin{array}{cc}
(\Im\Pi_{10})^2-\Im \Pi_{00}\Im \Pi_{20}
& \Im\Pi_{10}\Im\Pi_{01}-\Im \Pi_{00}\Im \Pi_{11}\\
\Im\Pi_{10}\Im\Pi_{01}-\Im \Pi_{00}\Im \Pi_{11} &
~~(\Im\Pi_{01})^2-\Im \Pi_{00}\Im \Pi_{02} \\
\end{array}
\right)\,
\ee
where the $\delta$-functions are written in terms of the imaginary part of the response functions $\Pi_{nm}(q,\omega)$ ($n,m=0,1,2$)
\be\label{EqPn}
\Pi_{nm}=4\sum_{\gamma,\gamma'}\sum_{\kv } \frac{F_{\kv ,\kv +\qv  }^{\gamma\gamma'}(v_{\kv ,\gamma} -v_{\kv +\qv  ,\gamma'})^n(\tilde{\epsilon}_{k,\gamma}v_{\boldsymbol{k},\gamma} -\tilde{\epsilon}_{|k+q|,\gamma'}v_{\boldsymbol{k}+\boldsymbol{q},\gamma'})^m(f_{0\kv ,\gamma} -f_{0\kv +\qv  ,\gamma'})}{\epsilon_{\kv ,\gamma} -\epsilon_{\kv +\qv  ,\gamma'}+\hbar\omega+i0^{+}}\,.
\ee
Equations (\ref{ICC}) and (\ref{EqPn}) are respectively represented in a more compact forms of Eqs. (5) and (6) in the main text.
\section{B. Disorder collision moments }
\setcounter{equation}{0}
\renewcommand{\theequation}{B.\arabic{equation}}
The non-momentum-conserving disorder collision integral is given by,
\begin{equation}
\boldsymbol{I}_{D\kv ,\gamma}  =\frac{2\pi g}{\hbar}\sum_{\gamma '}\sum_{\boldsymbol{k'}}|U_{\kv -\boldsymbol{k'}}|^2 F_{\kv,\kv'}^{\gamma\gamma'}(f_{\kv ,\gamma } -f_{\kv ',\gamma '})\delta(\epsilon_{\kv ,\gamma }-\epsilon_{\kv ',\gamma '})
\end{equation}
where the factor $g=4$ accounts for the spin and valley degeneracy. Considering the short-range disorder
characterized by an effective strength of $|U_{\kv -\boldsymbol{k'}}|\approx (n_dV_0^2)^{1/2}$ ($n_d$ is the disorder density and $V_0\delta(\bf r)$ is the disorder potential), the linearized collision integral becomes

\be
\begin{split}
&\boldsymbol{I}_{D\kv ,\gamma}  =\frac{2\pi n_dV_0^2}{\hbar}\sum_{\gamma'}\sum_{\boldsymbol{k'}}
\left[\left(v_{\kv,\gamma}\frac{\partial f_{0\kv,\gamma}}{\partial \epsilon_{\kv,\gamma} }-v_{\kv',\gamma'}\frac{\partial f_{0\kv',\gamma'}}{\partial \epsilon_{\kv',\gamma'} }\right)\cdot\pv_n\right.+\nonumber\\
&\left.\beta\left(v_{\kv,\gamma}\tilde\epsilon_{\kv,\gamma}\frac{\partial f_{0\kv,\gamma}}{\partial \epsilon_{\kv,\gamma} }-v_{\kv',\gamma'}\tilde\epsilon_{\kv',\gamma'}\frac{\partial f_{0\kv',\gamma'}}{\partial \epsilon_{\kv',\gamma'} }\right)\cdot\pv_s\right] \delta(\epsilon_{\kv,\gamma}-\epsilon_{\kv',\gamma'})
\end{split}
\ee
in which we set $F_{\kv,\kv'}^{\gamma\gamma'}=1$ for the sake of simplicity and without loss of qualitative behavior of disorder. Using $\delta(\epsilon_{\kv,\gamma}-\epsilon_{\kv',\gamma'})=\delta(\kv-\kv')/\hbar |\vv_{\kv,\gamma}|$, we easily construct the moments of the disorder collision integral
\be\label{IdMoments}
\left(\begin{array}{c}\sum_{\kv,\gamma}\vv_{\kv,\gamma}\boldsymbol{I}_{D\kv ,\gamma} \\
\sum_{\kv,\gamma}\beta\tilde\epsilon_{\kv,\gamma}\vv_{\kv,\gamma}\boldsymbol{I}_{D\kv ,\gamma}
\end{array}\right)=\Iv_D \cdot\left(
  \begin{array}{c}
    \pv_n\\
  \pv_s
  \end{array}
\right),
\ee
where,
\be\label{Id}
\Iv_D=
\frac{n_dV_0^2}{2\hbar^2}\sum_{\kv,\gamma} \vv_{\kv,\gamma}\frac{\partial f_{0\kv,\gamma}}{\partial \epsilon_{\kv,\gamma} }\left(\begin{array}{cc}
1 & \beta\tilde\epsilon_{\kv,\gamma}\\
\beta\tilde\epsilon_{\kv,\gamma} & (\beta\tilde\epsilon_{\kv,\gamma})^2
\end{array}\right).
\ee
Taking $k_F$, $v$, and $\beta^{-1}$,  respectively as units of wave vector, velocity and energy, $\Iv_D$  can be further simplified to, 
\be\label{Id2}
\Iv_D=
\frac{\lambda\beta}{4\pi}\left(\frac{\varepsilon_F}{\hbar}\right)^3\sum_{\gamma}\int_0^\infty dk~k^2 \gamma|\vv_{\kv,\gamma}|
\frac{\exp(\tilde\epsilon_{\kv,\gamma})}{[1+\exp(\epsilon_{\kv,\gamma})]^2}
\left(\begin{array}{cc}
1 & \beta\tilde\epsilon_{\kv,\gamma}\\
\beta\tilde\epsilon_{\kv,\gamma} & (\beta\tilde\epsilon_{\kv,\gamma})^2
\end{array}\right),
\ee
where the dimensionless disorder strength $\lambda=n_dV_0^2/(\hbar v )^2$. We solve the integrals in Eq. (\ref{Id2}) numerically for the full energy bands of bilayer graphene. 

\end{widetext}

 \end{document}